\documentclass[aps,pre,showpacs,preprint]{revtex4}

\usepackage{epsfig,graphicx}% Include figure files

\usepackage{bm}% bold math

\usepackage{epsfig}

\usepackage{amsmath,amssymb,amsfonts,wasysym,color}

\def\be{\begin{equation}}

\def\ee{\end{equation}}

\def\bea{\begin{eqnarray}}

\def\eea{\end{eqnarray}}

\begin{document}

\title{Solvation forces in Ising films with long-range boundary fields: density-matrix renormalization-group study.}

\author{A. Drzewi\'nski,$^{1}$ A. Macio\l ek,$^{2,3,4}$   and  A. Barasi\'nski,$^{1}$}

\affiliation{$^{1}$Institute of Physics, University of Zielona G\'ora, ul. Prof. Z. Szafrana 4a,
65-516 Zielona G\' ora, Poland}

\affiliation{$^{2}$Max-Planck-Institut f{\"u}r Metallforschung, Heisenbergstr.~3, D-70569 Stuttgart, Germany}

\affiliation{$^{3}$Institut f{\"u}r Theoretische und Angewandte Physik,
  Universit{\"a}t Stuttgart, Pfaffenwaldring 57, D-70569 Stuttgart, Germany}

\affiliation{$^{4}$Institute of Physical Chemistry, Polish Academy of
  Sciences, Department III, Kasprzaka 44/52,  PL-01-224 Warsaw, Poland}

\begin{abstract}
Using the quasi-exact density-matrix renormalization-group method
we calculate the solvation forces 
in two-dimensional Ising  films  of thickness $L$
subject to identical algebraically decaying boundary fields with various 
decay exponents $p$. 
At the bulk critical point  the solvation force acquires a universal contribution
which is long-ranged in $L$ due to the critical fluctuations, a phenomenon known as 
the critical Casimir effect.  For $p=2, 3$ and $50$, we study the scaling behaviour
 of the solvation force along  the pseudo-phase coexistence and  along  the critical and  sub-critical isotherms.

\date{\today}

\pacs{05.50.+q, 68.35.Rh, 68.08.Bc}

\end{abstract}

\maketitle

\section{Introduction}
\label{sec:1}

The solvation force $f_{solv}$  is the force per unit area between two surfaces, 
large colloidal particles or macromolecules
due to the intervening fluid.
It arises in the thermodynamic description of confined fluids 
as an excess pressure over the bulk value (fixed by the reservoir) 
and is conjugate to the distance $L$  between the confining surfaces \cite{evans}. 
The study of this generalized force for simple and complex fluids
is useful as it  may play an important role  in the field of colloid sciences, 
in  self-assembling systems and  in the  biology of protein folding. 
From a fundamental viewpoint, solvation forces  present an abundance  
of interesting problems to investigate: the behaviour at the phase transitions of a confined fluid,
such as capillary condensation \cite{evans,gubbins,mdb,dsd},  bridging \cite{bbd,apv,sumh},
localization-delocalization transition \cite{parry_evans,maria}, or layering  \cite{maria}, 
the structure  at small  separations $L$ \cite{israelchvili,christenson},
or the dependence on the fluid-fluid interactions and the  surface potentials \cite{attard,mdb,dsd}.
An addition appeal lies in the fact that it  may
be measured experimentally by a number of different techniques \cite{israelchvili,christenson}; 
e.g., the surface force apparatus,  the atomic force microscope, 
or the total internal reflection microscopy (TIMR) \cite{TIMR}. 
By using the latter technique,  the first direct determination of the  universal contribution to 
$f_{solv}$ between a single colloidal particle  and a planar surface 
in the presence of a near-critical binary liquid mixture was provided \cite{nature,lv}.
The universal  contribution, called the critical Casimir force \cite{FdG,krech99,dantchev,revs1}, 
is induced by  the confinement of  critical fluctuations of  the
order parameter of the second order phase transition, e.g. the density of the fluid or the  concentration of
one species for a binary liquid mixture.
At the bulk critical point this fluctuation-induced force has a leading power
law decay $f_{solv}/(k_BT_c)\sim AL^{-d}$ as $L\to \infty$, 
where $d$ is the spatial dimension of the system
and $A$ is the Casimir amplitude.
The  finite-size scaling theory  predicts (see, e.g., Ref.~\cite{barber})
that in the vicinity of the bulk critical point  $f_{solv}$
is described by a scaling function that depends on the so-called universality
class of the phase transition occurring in the bulk and on the geometry 
and surface universality classes of the confining surfaces \cite{diehl}.
The  critical Casimir  force is a  subject of recent considerable 
theoretical and experimental interest  
because at the submicron  scale its strength and range 
is comparable with other interactions and because it can be easily controlled,
including its sign,  by changes of  thermodynamic fields, such as the temperature, 
and by appropriate surface treatments. 

As mentioned above, the solvation force is associated with 
fluid-fluid interactions and the surface potential.
In the present paper we want to study the relevant 
effects of long-ranged, i.e., algebraically decaying 
forces on the near-critical behaviour of  $f_{solv}$ between two parallel walls
and test   predictions of the critical finite-size scaling theory. 
Examples of such long-ranged forces are dispersion or van der Waals forces, 
dipolar forces, RKKY interactions, forces in charged systems, and elastic forces in solids.
Here we utilize the equivalence  between a lattice gas model of a fluid 
and the Ising model and consider  Ising spin films
subject to identical boundary fields decaying in the orthogonal direction at a distance
$j$ from the
surface as $-h_1j^{-p}$, with $p= 2, 3, 50$ and $h_1\ge 0$. Using the
density-matrix renormalization-group (DMRG) method, we calculate
the solvation force  in two-dimensional ($2D$) systems along various thermodynamic paths.
The DMRG method  is  based on the transfer matrix approach and provides 
a numerically very efficient  iterative truncation
algorithm for constructing the effective transfer matrices for  
strips of fixed width and infinite length. The advantage of this method  is 
that it can treat arbitrary fields that are coupled to the single spin
variable. Moreover, because the DMRG method gives  quasi-exact results for the spectrum of the 
transfer matrix of the system,  fluctuations  are fully accounted for in this approach.
Recall, that in $2D$ systems  fluctuation effects are particularly strong.
However, this method works only if the spin-spin interactions are nearest-neighbour and, at present, 
it is limited to the two-dimensional systems. 

The studies carried out here  complement and extend the earlier 
works (see Refs.~\cite{dme,mde,mdb}).
In Refs.~\cite{dme,mde} it was argued that the  solvation (Casimir)
force in the vicinity of the critical point is strongly 
influenced by capillary condensation, i.e., 
the shift of the bulk first-order transition which occurs below
the critical temperature $T_c$. The residual condensation leads to the solvation force 
which is  much more attractive at temperatures 
near $T_c$ and a reservoir densities slightly below the critical value 
(or compositions slightly away from the critical composition in a binary mixture) than the Casimir 
value (at the bulk critical point) for the same $L$.
These predictions were supported by explicit calculations for
$2D$ Ising films with {\it short-ranged} (contact) boundary fields. 
By using the DMRG method it was possible to study the region between the capillary and bulk
critical points, i.e., for nonvanishing ordering  field $h<0$, which corresponds to the chemical
potential difference $\mu-\mu_{sat}$. In particular, 
the scaling functions of $f_{solv}$  along several isotherms were obtained showing that upon 
increasing the temperature towards $T_c$ a weakly rounded in $2D$ discontinuous jump of $f_{solv}$ on crossing
the coexistence line  \cite{evans,evans_UMBM} transforms gradually into a minimum.
A location of this minimum follows roughly the continuation 
of the (pseudo) capillary condensation line towards $T_c$ ~\cite{dme}.
The similar behaviour was found in the field-theoretic model solved numerically  in the mean-field
approximation \cite{frank}, again for the short-ranged  boundary fields.

In Ref.~\cite{mdb}, the case of short-ranged fluid-fluid interaction and long-ranged wall-fluid potentials
decaying as $-Az^{-p}$ for $z\to \infty$ was considered for  $2D$ Ising films  and for the truncated
Lennard-Jones fluid in a slit geometry. The discrete model was treated within the DMRG method whereas
the continuum one by the nonlocal density functional theory.
The study was focused on the asymptotic behaviour  of the solvation force as $L\to \infty$.
Except for a high temperature Ising system, results for both models  agree with the predictions from
the analysis based on the wall-particle Ornstein-Zernike equations that  $f_{solv}$ is repulsive and
asymptotically decays with the same power law as the wall-fluid potential; this prediction holds away
from the critical temperature and from any phase transition. For  Ising  films
above $T_c$ the asymptotic behaviour was found to be  of the  higher order then that of the boundary field,
i.e., $~L^{-(p+1)}$. This was explain by the specific symmetry of the order parameter 
with the  spontaneous magnetization  $m^*$ equal to zero above $T_c$.
Moreover, $f_{solv}$ was calculated along the {\it  bulk} two-phase coexistence line slightly
on the liquid side of this line and  at the critical density 
for $T>T_c$; in the Ising system this path corresponds to the line $H=0$. The obtained results 
imply that for ordering field $H=0$ and $L\to \infty$
\begin{equation}
\label{eq:1}
f_{solv}\sim f^{reg}_{solv}+f^{sing}_{solv}=2\rho BL^{-p}+(d-1)L^{-d}\vartheta(L/\xi_{\tau}),
\end{equation}
where $\rho$ is the bulk density and  $B$ is related to the strength of the wall-fluid potential. For Ising
systems $\rho B$ is replaced by $m^* h_1$.  $\vartheta(x)$ is a universal scaling function describing the 
contribution arising from  critical fluctuations of a fluid. $\tau =(T-T_c)/T_c$ and  $\nu$ is the critical
exponent of the  bulk correlation length $\xi_b$.  $\vartheta$  is vanishingly small away from the critical
region and is negative for identical walls.

In the present study we consider the neighbourhood of  the bulk critical point and
investigate the scaling behaviour of the solvation force along the critical and two subcritical
isotherms and along the pseudo-coexistence (capillary condensation) 
line slightly on the "liquid" and on the "gas" sides. A substantial progress in computer capacities allows
to study sufficiently thick films to assure that the scaling limit is achieved. Here we consider strips of
widths up to $L=700$ lattice constants. According to  general scaling arguments \cite{diehl,dsd} the finite-size
behaviour of the singular part of the solvation force is modified by the presence of the long-ranged
substrate-wall potentials. However, if $p>(d+2-\eta)/2$ the universal behavior is expected to
hold. In this case the long-ranged part of the boundary field is irrelevant in the RG sense with respect to 
a pure contact surface field \cite{diehl}. Here $\eta$ is the  critical exponent governing 
the algebraic decay of the two-point correlation function in the bulk and at $T_c$.
In the present case  of the $d=2$ Ising model $\eta(d=2)=1/4$ so that 
for $p> 15/8$ we expect to observe  the power law $L^{-2}$.
The scaling of the solvation force in films with long-ranged fluid-fluid 
and substrate-fluid potentials
was analysed in Ref.~\cite{dsd} by using the general scaling arguments 
and mean-field theory. We will summarize the relevant conclusions of this analyses
and relate our results to them in Sec.~\ref{sec:3}.

Our paper is organized as follows. In Sec.~\ref{sec:2} we introduce the model and describe 
the determination of the phase diagrams for various values of the parameter $p$.
Sec.~\ref{sec:3} contains our results. Finally, we give conclusions in Sec.~\ref{sec:4}. 

\section{Microscopic model}
\label{sec:2}

We consider $D=2$ Ising  strips defined on a square lattice of size $M\times L, M\to \infty$ 
and  subject to the same  boundary fields on both sides. The lattice consists of $L$
parallel rows at spacing $a$, so that the width of the strip is $La$;
in the following we set $a=1$. At each site there is an Ising 
spin variable taking the value $\sigma_{k,j}=\pm 1$, where $(k,j)$ labels the site.
The boundary surfaces  are located in the rows  $j=1$ and $j=L$ and periodic boundary conditions 
(PBCs) are assumed in the lateral  $x$ direction. The Hamiltonian of our model is given by
\begin{eqnarray}
\label{eq:2}
{\cal H} & = & -J\left( \sum_{\langle kj,k'j'\rangle}\sigma_{k,j}\sigma_{k',j'}\right. \nonumber \\
\mbox{}&\mbox{}&+\left.\sum_{j=1}^{L}V^{ext}_{j,L}\sum_k\sigma_{k,j}+H\sum_{k,j}\sigma_{k,j}\right),
\end{eqnarray}
where the first sum is over all nearest-neighbor pairs and
the external potential is measured in units of $J>0$.
$V^{ext}_{j,L}=V^{s}_{j}+V^{s}_{L+1-j}$ is the total boundary field experienced by a spin in 
row $j$; it  is the sum of the two independent wall contributions.
The {\it s}ingle-boundary field $V^s_{j}$ is taken  to have the  form
\begin{equation}
\label{eq:V_j}
V^s_{j}=\frac{h_1}{j^p}
\end{equation}
with $p>0$ and $h_1>0$. $H$ is a bulk magnetic field. 
$h_1$ and $H$ are dimensionless (see Eq.~(\ref{eq:1})). 

As already mentioned this model is equivalent to  
the $2D$ lattice gas model of a two-dimensional one-component  fluid
with a short-ranged interaction potential between the fluid particles and either short-ranged
or long-ranged substrate potentials (see, e.g., Ref.~\cite{pandit}).

\subsection{Phase diagram}
\label{sec:pd}

\begin{figure}[hbt]
\centering
\includegraphics[scale=0.55]{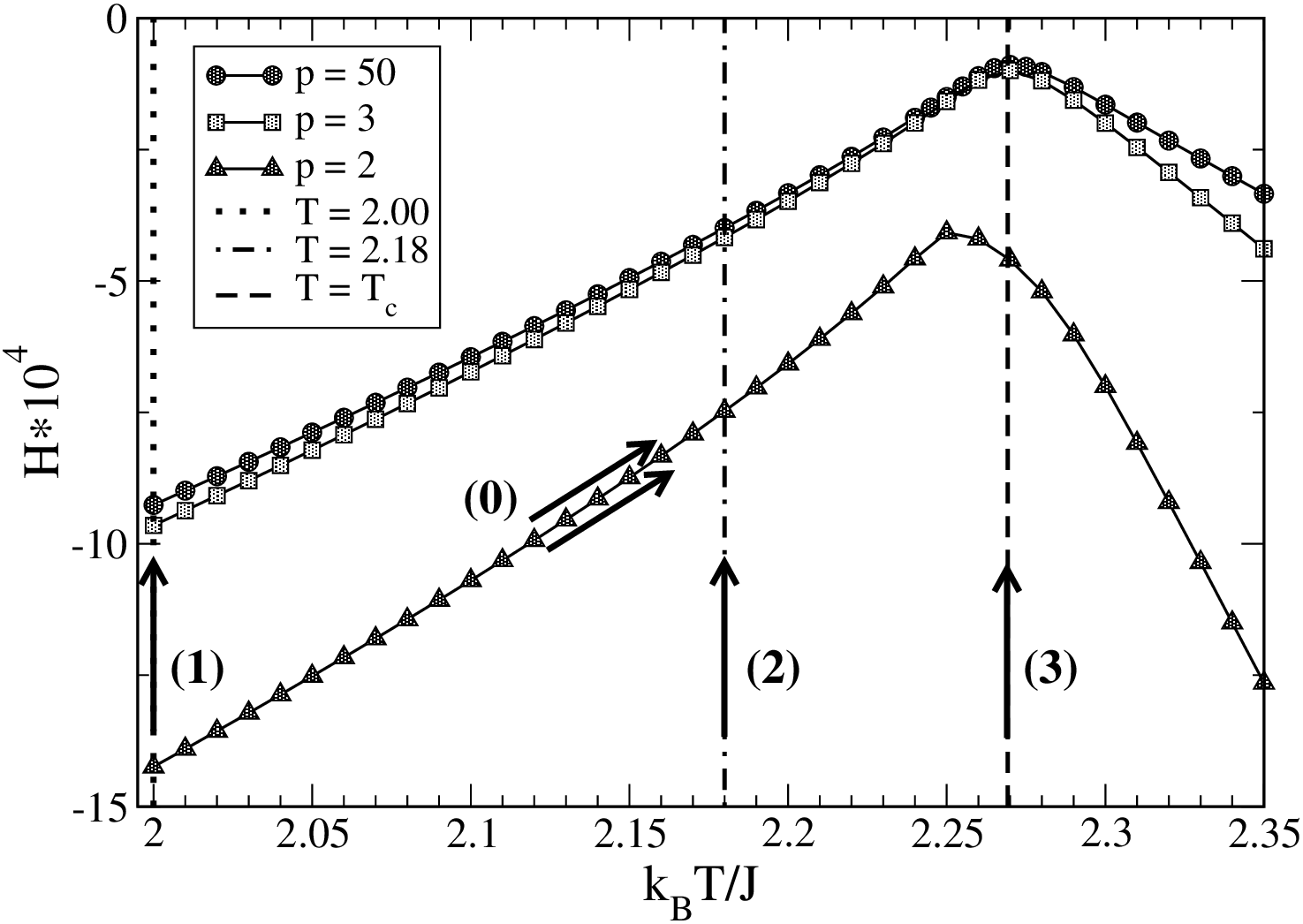}
\caption{Phase diagram for a  $2D$ Ising strip subject to identical
boundary fields  $V^{ext}_{j}$ (see Eq.~(\ref{eq:V_j})) obtained by using
DMRG for a strip width  $L=600$ and the amplitude $h_1$
for the boundary fields such that $x=h_1*L^2=20000$. The thick solid line indicates the 
bulk coexistence line at $H=0$ ending at the bulk critical point (black circle). 
The lines interpolating the symbols represent the pseudo-phase coexistence lines $H_{ca}(T,L;p)$ 
for three different values of the exponent $p$ governing the algebraic decay 
of the boundary fields. Three isotherms along which the solvation force  has been calculated are shown - paths
$(1)$-$(3)$. The other thermodynamic path $(4)$ and $(5)$ run parallel to the pseudo-coexistence line
slightly on the "liquid" and on the "gas" side, respectively. 
}
\label{fig:1}
\end{figure}

In Fig.~\ref{fig:1} we show the phase diagram for the present model
calculated by using the DMRG method for a strip of  width 600 and for three
choices of the parameter $p$ describing the decay of the boundary field: 
$p=2, 3$, and 50. $h_1$ is chosen  so that the scaling variable $x=h_1*L^2=20000$, which 
for the short-ranged boundary fields is sufficient to ensure each system corresponds to the infinite surface
field scaling limit \cite{DMC:2000}. The case  $p=50$ is expected to resemble the behavior 
corresponding  to short-ranged surface forces. In this figure we  display the various thermodynamic paths
along which we have calculated the solvation force.

For surfaces which  prefer the same
bulk  phase, the phenomenon equivalent to capillary condensation takes place.
The pseudo-phase coexistence between phases of spin up 
and spin down occurs along the line $H_{ca}(T,L;p)$, 
which is given approximately by the analogue of the Kelvin equation  \cite{kelvin}.
For positive surface fields capillary condensation occurs at negative values $H_{ca}(T,L;p)$
of the  bulk field $H$. The pseudo-coexistence lines   have been identified as those  positions $(H,T)$
in the phase diagram where the total magnetization of the strip vanishes, i.e., $\sum_{j=1}^Lm_{j}=0$
with $m_{j}=\langle \sigma_{k,j} \rangle$. We have not attempted to localize the position of the 
pseudo-critical temperature $T_{c,L}$  which ends the pseudo-coexistence lines; the unique  determination 
of the pseudo-critical point is not possible because non-analytic behaviour is rounded in $2D$ \cite{cross}.
Notice, that for $p=3$ and 50 and $T>T_c$,  the line defined  by the zeros of the total magnetization moves
to more negative values of $H$ upon increasing temperatures. For $p=2$ we observe this trend already below $T_c$. 
In Ref.~\cite{cross} the shift of   pseudo-phase coexistence lines $H_{ca}(T,L;p)$ for $p=2$ and 3 relative to
the  short-ranged pseudo-phase coexistence line ($p=50$) was analysed. For $p=50$ and 3 also the scaling behaviour 
of the pseudo-coexistence line of capillary condensation was studied.

The thick solid  line  in  Fig.~\ref{fig:1} indicates the bulk phase coexistence line $(H=0, T<T_c)$
terminating at the bulk critical point  $(H=0,T=T_c\simeq 2.269 J/k_B)$ (black circle). 
The  symbols (triangles for $p=2$, squares 
for $p=3$, and circles  for $p=50$) show the pseudo-phase coexistence.
It turns out  that the pseudo-phase coexistence line
for long-ranged boundary fields is located slightly further 
away from the bulk coexistence line, especially at  lower temperatures,
than the pseudo-phase coexistence line for the short-ranged boundary fields ($p=50$).

\section{Results}
\label{sec:3}

Before presenting our results let us summarize predictions for the  scaling 
behaviour of the solvation force based on general scaling arguments \cite{dsd}.
One expects that the finite-size behaviour of the singular part of the solvation force 
is modified due to the presence of long-ranged substrate-fluid potentials:
\begin{equation}
\label{eq:3}
f_{solv} L^{d}\simeq \vartheta\left[L/\xi_{\tau},L/\xi_{H},
\left(L/\xi_0\right)^{-\omega_s}h_1,\left(L/\xi_0\right)^{-\omega}g_{\omega}\right]
\end{equation} 
Here  $\xi_{\tau}(\tau\to\pm 0,H=0)=\xi_0^{\pm}|\tau|^{-\nu}$ is the bulk correlation length at bulk 
coexistence $H=0$ while  $\xi_H(\tau=0,H)=\xi_{0,H}|H|^{-\nu/\Delta}$ is the
bulk correlation length at the critical temperature $T=T_c$. $\omega_s=p-(d+2-\eta)/2$ is the correction to
scaling exponents due to the long-ranged tail of the substrate-fluid
interactions. $\omega$ is the Wegner's \cite{wegner} correction-to-scaling exponent for short-ranged systems, 
which is equal to $1/2$ for $2D$ Ising model. $g_{\omega}$ 
is a dimensionless nonuniversal  scaling field. In the case of $p=50$ that mimics the effect of the contact surface
fields, the third argument of the scaling function in (\ref{eq:3}) should be replaced by $Lh_1^{\nu/\Delta_1}$,
where the surface gap exponent $\Delta_1=1/2$ for $2D$ Ising system.
For large $L$ and $\omega_s>0$, one can expand the scaling function $\vartheta$ in (\ref{eq:3})
\begin{eqnarray}
\label{eq:4}
f_{solv}L^{d} \simeq &&
\vartheta^{sr}\left[L/\xi_{\tau},L/\xi_{H}\right]+
\left(L/\xi_0\right)^{-\omega_s}h_1\vartheta^{lr}_s\left[L/\xi_{\tau},L/\xi_{H}\right]\\ \nonumber
&  +&
\left(L/\xi_0\right)^{-\omega}g_{\omega}\vartheta^{sr}_{\omega}\left[L/\xi_{\tau},L/\xi_{H}\right].
\end{eqnarray} 
Thus for $L\simeq \xi_{\tau},\xi_H$,  $\vartheta^{lr}_s$ and $\vartheta^{sr}_{\omega}$ represent corrections to 
 the leading $L$-dependence provided by $\vartheta^{sr}$. For $L/\xi \gg 1$
the scaling function $\vartheta^{sr}$ decays exponentially, therefore in this regime $\vartheta^{lr}_s$ 
is the leading finite-size contribution in the singular behaviour of the force.
As mentioned already in the Introduction, in the present case  of the $2D$ Ising model $\eta=1/4$ so that
the expansion (\ref{eq:4}) should hold  for $p> 15/8$, i.e., for all  cases considered in our work.

However, one can argue that for sufficiently thin films the contribution of the long-ranged substrate-fluid
potential is always important, also at the bulk critical point.
Because the separation between two surfaces  is finite and the substrate-fluid  potentials are 
long-ranged and identical, there will be always  some nonzero field acting at the center of the film.
This effect can be interpreted as if the system experiences an external bulk  field 
$H_{eff}=2h_1\left[L/(2\xi_0)\right]^{-p}$, although  in fact
it might be at the bulk coexistence curve $H=0$. The relevance of 
finite-size contributions due to this effective bulk field can be estimated by taking into account that
$H$ scales as $HL^{\Delta/\nu}$. These finite-size contributions are negligible in the critical regime if
$2^{p+1}|h_1|\left[L/\xi_0\right]^{\Delta/\nu-p}\ll 1$. From that inequality one can identify the
thickness of the film $L_{crit}$ such that for $L\le L_{crit}$ the effect due to the long-ranged
substrate-fluid potentials is  relevant   $L_{crit}\simeq \xi_0\left(2^{p+1}|h_1|\right)^{-1/(d-p-\beta/\nu)}$.
For $2D$ Ising model 
$\beta/\nu=1/8$ and $\xi_0 \approx 0.5673$ therefore for $p=2$ one obtains
$L_{crit}\simeq  0.5673 \left(8|h_1|\right)^{8}$, i.e.,  the long-ranged tails practically always matter.
For $p=3$,  $L_{crit}\simeq  0.5673 \left(16|h_1|\right)^{8/9}\simeq 7\left(|h_1|\right)^{8/9}$
which assuming $|h_1|=10$ gives $L_{crit} \approx 50$. The separation $L$ is in units of a lattice constant.

The total excess free energy per unit area for the case of identical surface fields $h_1=h_2$ and
non-vanishing bulk magnetic field $H$ can be written as
\begin{equation}
\label{eq:exfree}
f_{ex}(L)\equiv L(f(L,T,H,h_1)-f_b(T))+2f_w(T,H,h_1)+f^*(L,T,H,h_1)
\end{equation}
where $f$ is the free energy per site, $f_b$ is the bulk free energy, $f_w$ is the $L$-independent
surface excess free energy contributed from each wall, and $f^*$ is the finite-size contribution to
the free energy. All energies are measured in units of $J$ and the temperature in units of $J/k_B$.
$f^*$, which vanishes for $L\to \infty$, gives rise to the generalized force, which is analogous to
the solvation force between the walls in the case of confined fluids,
\begin{equation}
\label{eq:solfordef}
f_{solv}=-(\partial f_{ex}(L)/\partial L)_{T,H,h_1}.
\end{equation} 

In the transfer matrix approach the leading eigenvalue $\lambda_L$ of the transfer matrix $T_L$ 
\begin{equation}
\label{eq:dmrg1}
T_L |v_L\rangle = \lambda_L |v_L\rangle,
\end{equation}
gives the free energy per spin of an Ising strip as
\begin{equation}
\label{eq:dmrg2}
\beta f(L)=-\frac{1}{L} \ln \lambda_L .
\end{equation}
The components of the eigenvector $|v_L>$ related to the leading eigenvalue give
the probabilities of various configurations. In order to calculate the size $L$ dependence of
the solvation force at fixed values of parameters ($T,H,h_1$) we calculate the excess free energy per
unit area $ f_{ex}(L)\equiv \left(f-f_b\right)L$ at $L_0+1$ and $L_0-1$. Having values $f^{ex}(L_0+1)$ and
$f^{ex}(L_0-1)$ we approximate the derivative in eq.(~\ref{eq:solfordef}) by a finite difference
$f_{solv}=- (1/2)(f^{ex}(L_0+1)-f^{ex}(L_0-1))$. 

The case with  vanishing external field $H$ is
relatively convenient to study because for the bulk free energy there exists  an exact solution 
by Onsager \cite{onsager}. For non-zero $H$, in order to find the bulk free energy 
for each thermodynamic state one has to perform the calculation 
for finite systems and extrapolate the data to $L \to \infty$. We have determined the largest eigenvalue
of $T_L$ for the strips ($L \times \infty$) with free boundary conditions and widths $L$ ranging
from $380$ to $680$. Next the free energy was extrapolated to $L \to \infty$ by means of the Bulirsch and
Stoer method \cite{bulirsch}.

\subsection{Thermodynamic path along the pseudo -- phase coexistence  line}
\label{subsec:pseud}

\begin{figure}[hbt]
\centering
\includegraphics[scale=0.5]{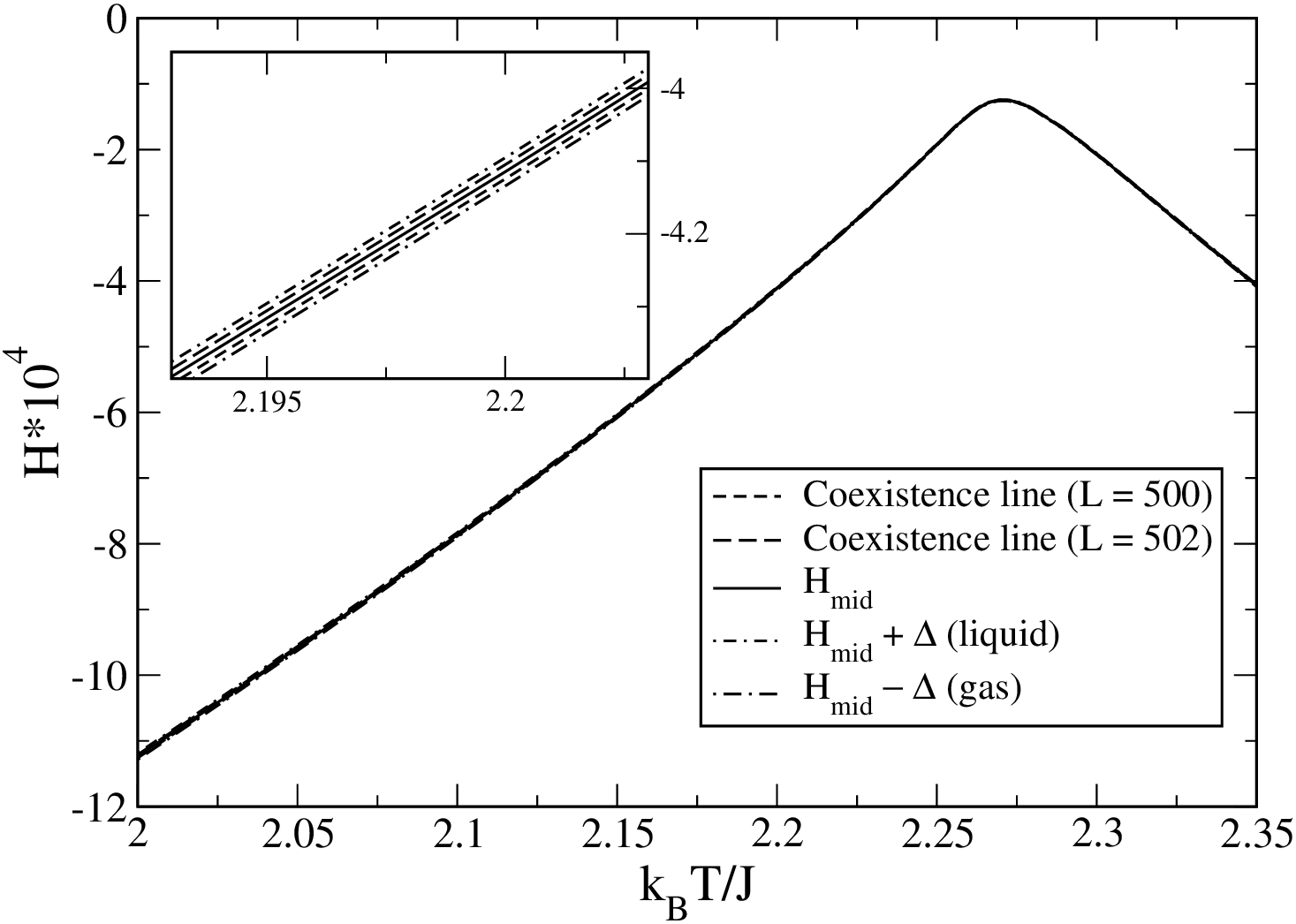}
\caption{Phase diagram for an Ising strip subject to identical short-ranged ($p=50$)
boundary fields $h_1\sim6.32$ for strip widths $L=500$ and $L=502$ corresponding to $x=h_1L^2=20000$.
The solid line denotes the middle line between  both pseudo-coexistence lines defined as
$H_{mid}(T) = \left( H_{co}(T;L=502)+H_{co}(T;L=500) \right)/2$. Dashed lines indicate
the pseudo-coexistence lines, whereas the dashed-dot lines present the thermodynamic paths shifted
with respect to the $H_{mid}$ line by $\Delta(T)=H_{co}(T;L=502)-H_{co}(T;L=500)$.
}
\label{fig:2}
\end{figure}

A new
computational problem arises when one wants to determine the solvation force  on the thermodynamic path along
the pseudo-coexistence line. In order to calculate the derivative of the excess free energy at certain $L_0$
the values of the
free energy are necessary at $L_0+1$ and $L_0-1$. According to the Kelvin equation, the pseudo-coexistence
is shifted from the bulk position $H=0$ proportionally to $1/L$. Therefore the path $(4)$ or $(5)$  cannot be
chosen to lie
close to the pseudo-coexistence line corresponding to $L_0$ (see Fig.~\ref{fig:2}). Our way of defining
the middle line between
the $L_0+1$ and $L_0-1$ pseudo-coexistence lines and the shift  $\Delta$ quarantees that
for both $L_0+1$ and $L_0-1$ one is  on the same (liquid or gas) side of the above coexistence lines.

\begin{figure}[hbt]
%\centering
\includegraphics[scale=0.6]{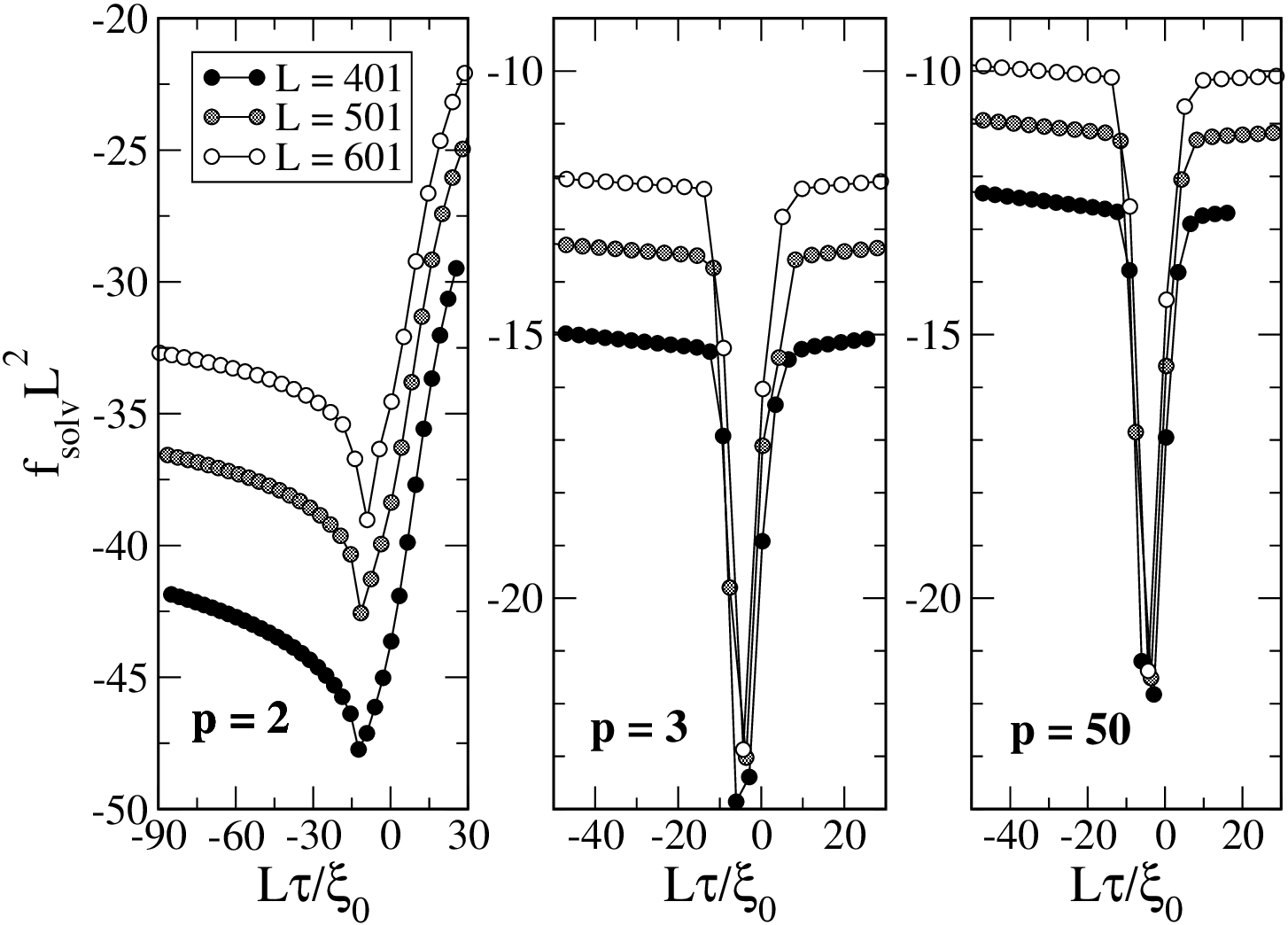}

\caption{Scaled solvation force as a function of $L/\xi_{\tau}=L\tau/\xi_0$ calculated for different separations
$L$ along the line of pseudo-coexistence 
slightly on its gas side - the path $(4)$. The amplitude of the substrate potential $h_1$ is chosen such that 
$Lh_1^2=20 000$. Different panels show results for different values of the exponent $p$ describing the 
algebraic decay of the substrate fluid potential. Scaling does not hold.}
\label{fig:3}
\end{figure}

\begin{figure}[hbt]
\centering
\includegraphics[scale=0.6]{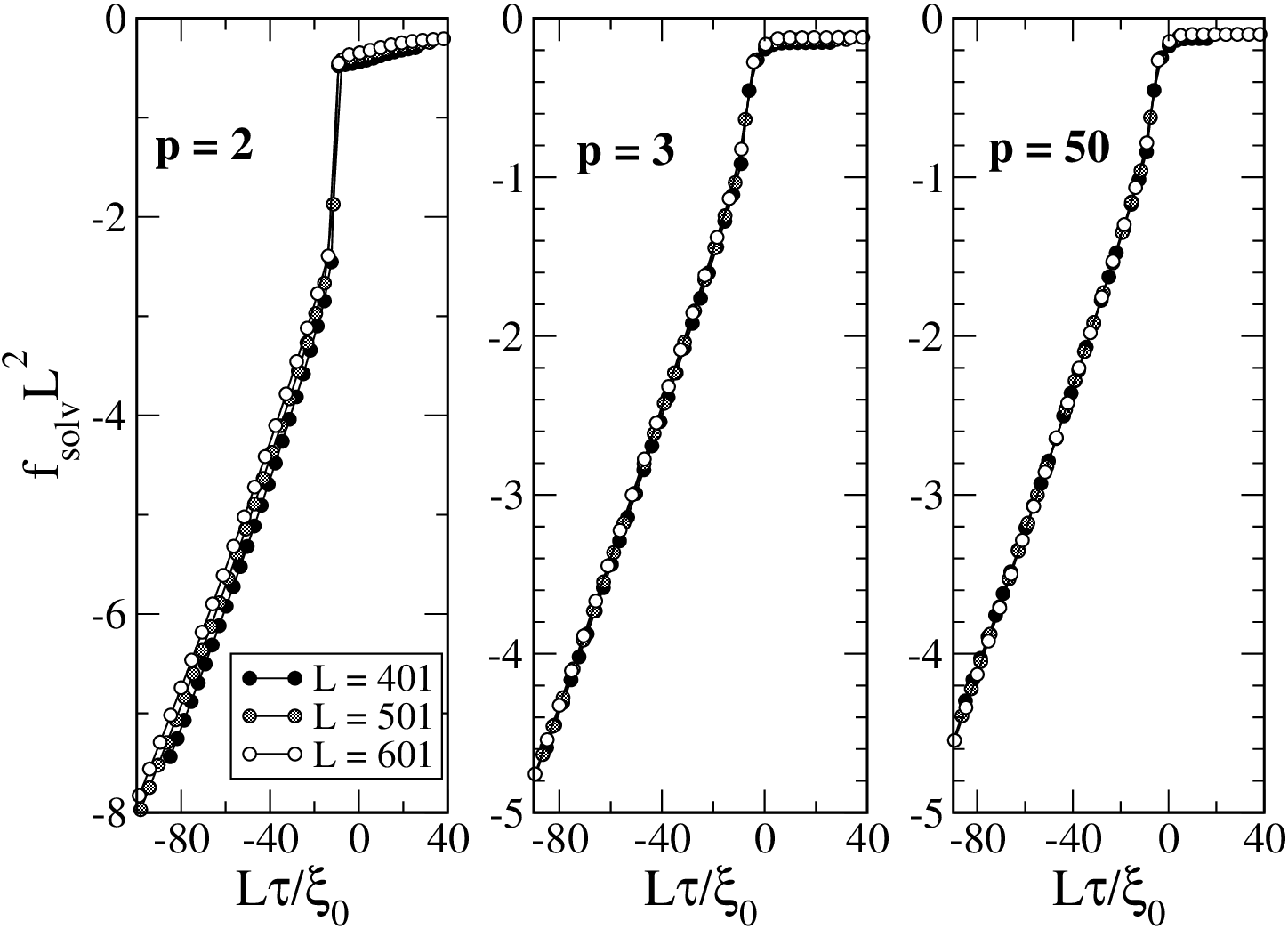}
\caption{The same as in Fig. 3 but slightly on  the liquid side of the pseudo-coexistence curve - the path $(5)$.
Scaling is very well obeyed.}
\label{fig:4}
\end{figure}

The solvation force scaled by $L^2$ as a function of the scaling variable $L/\xi_{\tau}$, evaluated along 
the pseudo-coexistence line  slightly on its gas side (path (4)) and slightly on its liquid side (path (5))
is shown in Fig.~\ref{fig:3} and Fig.~\ref{fig:4}, respectively. Data correspond to the fixed $x=Lh_1^2=20 000$
and different values of $L$. Note that in the case of $p=50$ which mimics a short-ranged contact surface field,
$x$ is the relevant scaling variable.
Panels (a), (b) and (c) correspond to $p=2, 3$, and 50, as depicted.
On both sides of the pseudo-coexistence line and for all values of the parameter $p$,
the force is attractive in the whole range of the studied scaling variable $L/\xi_{\tau}$. 

On the gas side, data corresponding to different separations $L$ do not collapse into a common curve, i.e.,
there is no scaling except for $p=3$ and 50 near the  minimum close to $T_c$. Moreover, the solvation force 
$f_{solv}^-$ is very weak - the weaker the larger value of the decay exponent $p$.
For $p=3$ and 50 the behaviour is qualitatively the same, we observe a slight linear variation of $f_{solv}^-$
with a "dip" in the narrow interval below but close to $T_c$. On the contrary, 
on the liquid  side of the pseudo-coexistence line an excellent scaling has been found  for $p=3$ and $50$.
For $p=2$ one can see deviations from scaling, especially below $T_c$.
The solvation force $f_{solv}^+$ is  much stronger than on the gas side - data shown 
in Fig.~\ref{fig:4}  correspond  to the scaling function divided by 100. The presence of the pseudo-capillary
critical point manifests itself as a crossover between  a  rapid linear increase of the solvation force at
low temperatures and a saturation at small values for  high temperatures.
In Ref.~\cite{mde} it was argued that at the transition slightly on the liquid side 
 $f^+_{solv}\approx 2H_{co}(T)m^*(T)\approx -2\sigma(T)/L$. It follows that  
the amplitude of $f^+_{solv}$ should increase in the same fashion as the interfacial tension as $T$ increases
at fixed $L$. For the $2D$ Ising model, the surface tension is given exactly by
$\beta J \sigma(T)=2(K-K^*)$, where $K=J/k_BT$,
$K^*= {\rm arcth} (\exp(-2K))$, and $K_c=J/(k_BT_c)=\ln(1+\sqrt 2)$ \cite{onsager}. Since
$\sigma(T) \simeq -4K_c\tau$, the scaling function should vary linearly with $\tau$ at fixed $L$, which agrees
with our result for $p=50$ and 3. The argument given in Ref.~~\cite{mde} is based on the macroscopic approximation
for the total free energy which ignores interactions between surfaces, i.e., it is valid for $L\to \infty$. 
For $p=2$  such interactions cannot be ignored, even for large $L$. Therefore it is not surprising that we
can see some  deviations from the linear variation of $f_{solv}^+$ in that case.
For the gas  side of the pseudo-coexistence curve, 
the analysis  in Ref.~\cite{mde} gives $f_{solv}^-\approx 0$. What we have found, is  a nonzero but very weak 
solvation force. It  arises  from the $L$-dependent part of the total free energy which 
is neglected in the previous argumentation. As mentioned in the Introduction, away from the bulk critical point
the dominant contribution to the solvation force comes from the regular term which decays with $L$ in the same way
as the substrate fluid potential, i.e., $\sim L^{-p}$ or  exponentially in $L$ for the case of short-range
substrate potential - this explains the observed  lack of the scaling behaviour.
Finally, we note that on  both sides of the pseudo-coexistence line $f_{solv}$ is attractive for all values
of $p$. In contrast, in the same range of temperatures along  the bulk coexistence line $H=0$ the solvation
force is repulsive for $p=2$ and 3 and attractive for short-range boundary fields \cite{mdb}.

\vspace{2cm}

\vspace{1cm}

\subsection{Isotherms}
\label{subsec:iso}

\begin{figure}[hbt]
\centering
\includegraphics[scale=0.6]{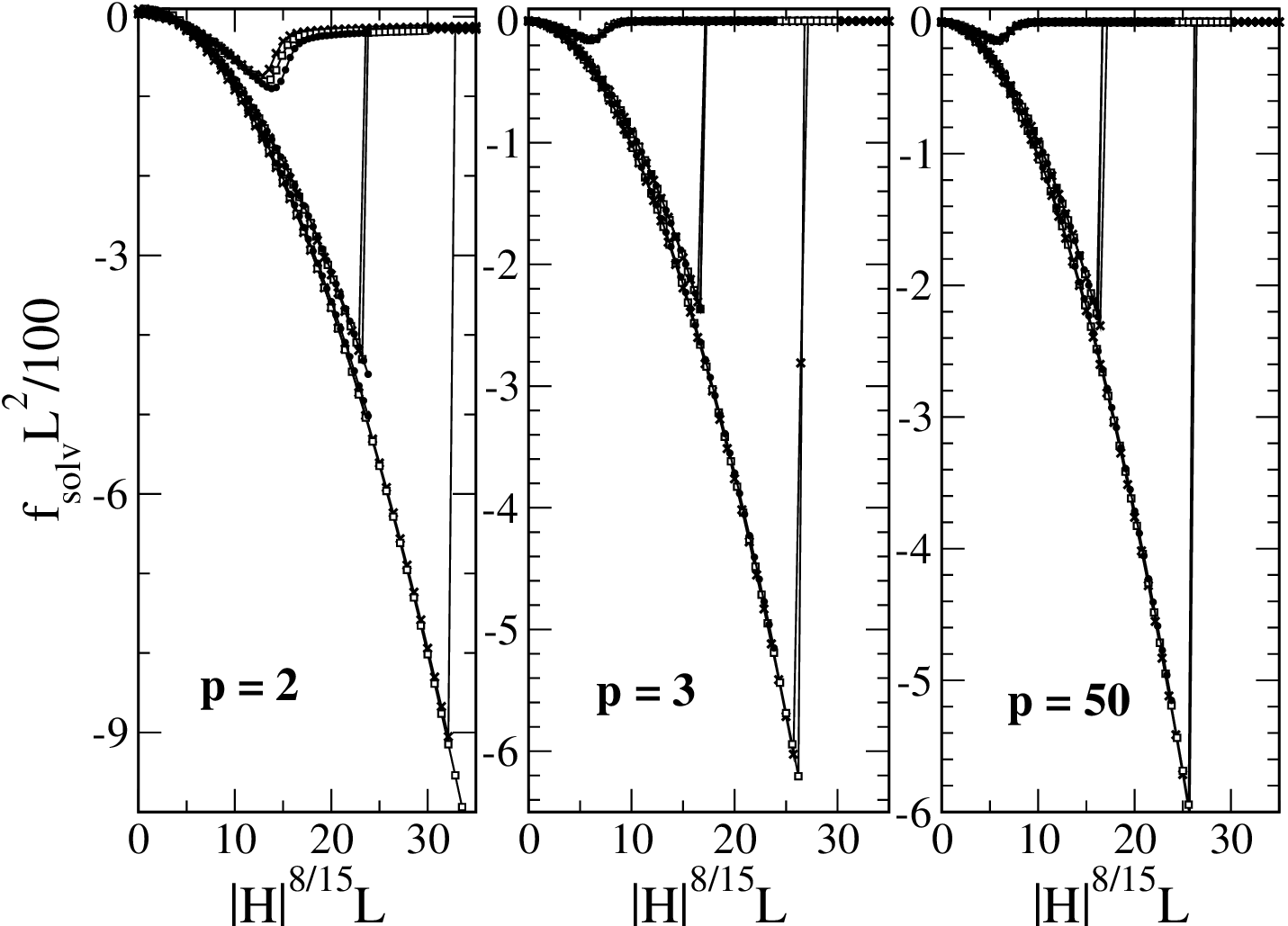}
\caption{Scaled solvation force as a function of $|H|^{8/15}L$ calculated for different separations $L$  along
3 various isotherms. For $L/\xi_{\tau} \approx 127$  (path (1)- bottom curve) and 42 (path (2)- middle curve), 
the sharp jumps to a strongly attractive solvation force as $|H|^{8/15}L$ is reduced indicate weakly rounded
capillary condensation of the $(+)$ (liquid) phase. Data corresponding to $L=601$ (crosses),  501 (squares),
and 401 (circles) follow common curves to the left from
the jumps. Along the critical isotherm solvation force exhibits a minimum. 
}
\label{fig:5}
\end{figure}

As mentioned in the Introduction, the scaling behaviour of the solvation force along various isotherms 
was studied for $2D$ Ising model with short-range contact boundary fields in Ref.~\cite{mde}.
The conclusion was that in the range of the scaling variable $L/\xi_{\tau}$ in which the solvation force
displays the 
features of (weakly rounded) capillary condensation, i.e., the jump from values appropriate to a spin down 
$(-)$ (gas) phase $f^-_{solv} \approx 0$ to the negative values appropriate to a spin up $(+)$ (liquid) phase
$f^+_{solv} \approx 2Hm^*(T)$, scaling is not well obeyed. 
In the present study we have been  able to perform calculations for  much bigger systems than
that considered in \cite{mde}. We find that along the isotherms corresponding to $L/\xi_{\tau}\approx 127$
and $\approx 42$,
data for different values of the separation $L$ collapse onto the common curve on the liquid side 
of the pseudo-coexistence line - see Fig.~\ref{fig:5}. 
This holds for all considered values of the parameter $p$. This is because on the liquid
side of the pseudo-coexistence line the leading behaviour of the solvation force $f^+_{solv} \approx 2Hm^*(T)$,
so that $f_{solv}L^2$ is approximately a quadratic function of  $H^{8/15}L$.
On the gas side, the leading behaviour is the regular part of the solvation force (see Eq.~(\ref{eq:1})).
The maximum absolute value of the force is 
$~\Delta f_{solv}= f^+_{solv}-f^-_{solv}\approx 2H_{co}(T)m^*(T)\approx -2\sigma(T)/L$,
which for a  fixed temperature decreases as $1/L$.
The location of the abrupt change in the value of the solvation force depends slightly on $L$, as it should be
for a weakly rounded transition.  What is also characteristic is that on the liquid side 
of the pseudo-coexistence curve the scaling function varies only very little from one isotherm to another.
As  $L/\xi_{\tau}$ decreases the jump of $f_{solv}$ gradually transforms into a minimum.

\begin{figure}[hbt]
\centering
\includegraphics[scale=0.6]{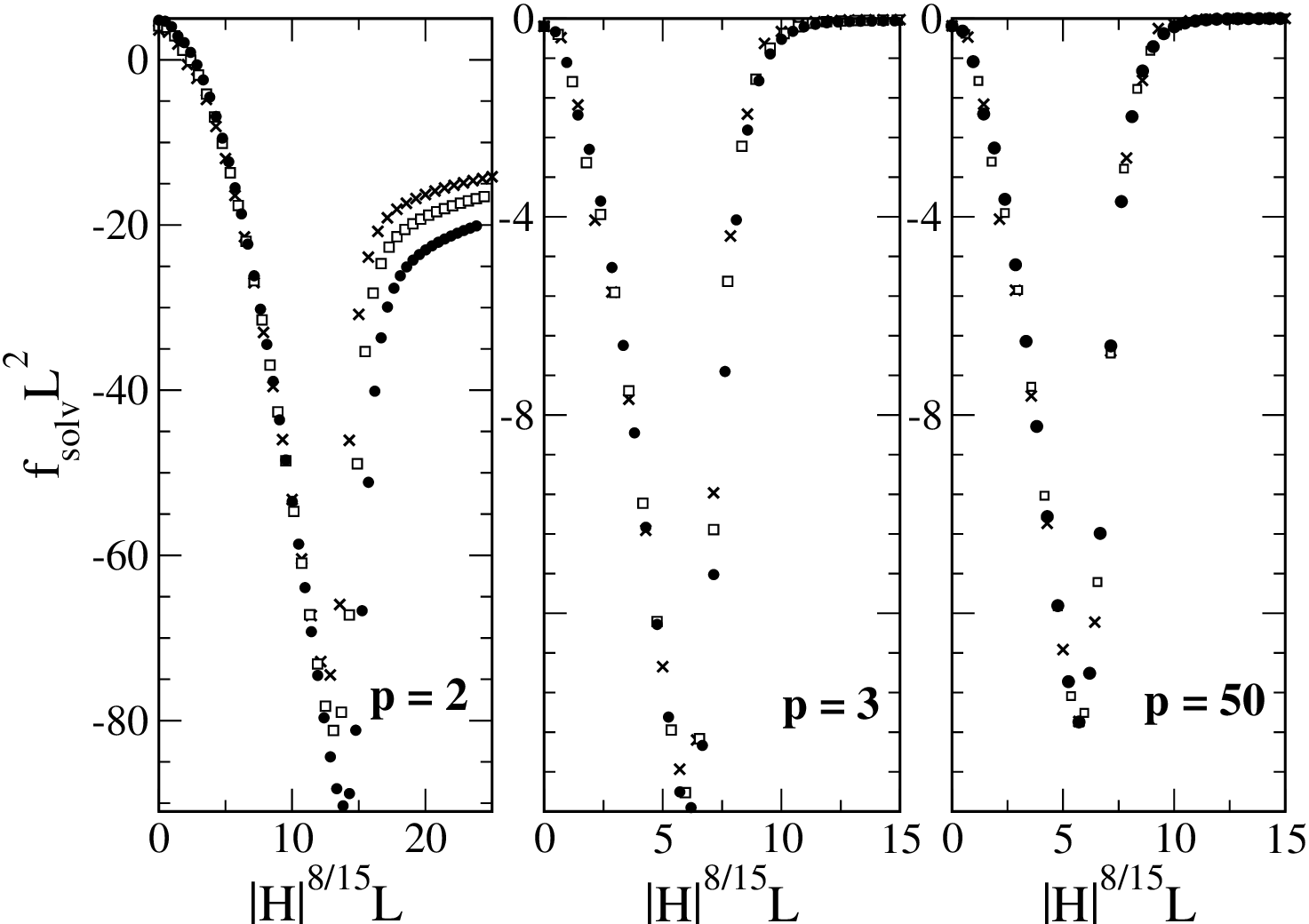}
\caption{Scaled solvation force as a function of $|H|^{8/15}L$ calculated along the critical isotherm.
 Scaling is excellent for $p=50$. For $p=3$ and 2 data collapse on the left shoulder of the minimum - to
the right from the minimum we observe 
deviations from scaling. For $p=2$ and 3, $f_{solv}$ becomes repulsive for small
values of $|H|^{8/15}L$.  Symbols have the same meaning as in the Fig.~\ref{fig:5}.
}
\label{fig:6}
\end{figure}

For  $p=50$, scaling along the critical isotherm is excellent  
in the whole range of the scaling variable $|H|^{8/15}L$  (see Fig.~\ref{fig:6}).
For $p=2$ and 3,  data collapse only
on the left shoulder of the minimum, which indicates some residual effect of the condensed phase.
As one can expect the finite size effects increase with the range of the substrate potential.
Notice also that for long-ranged boundary fields the maximal absolute value of the solvation force is strongly
increased  with respect to the one for the short-ranged one and that the force becomes repulsive for 
$|H|^{8/15}L \to 0$. 

\section{Conclusions}
\label{sec:4}

We have presented  DMRG results for the solvation
force $f_{solv}$ for $2D$ Ising systems between identical walls with  surface potentials 
decaying algebraically with the exponent $p=2, 3$ and 50.
The range of the substrate potential does not influence the qualitative
behaviour of the solvation force.
We found, in agreement with simple macroscopic arguments, that on the liquid side of the pseudo-coexistence
line the behaviour of the solvation
force is dominated by the  bulk excess free energy associated with the  $(+)$ phase
being metastable in the  bulk. Thus $f^+_{solv}$ is almost $L$-independent along the 
isotherms (for temperature sufficiently below $T_c$) which leads to the data collapse 
in plots of $f_{solv}L^2$ vs $|H|^{8/15}L$. Along
the pseudo-coexistence line,   $f^+_{solv}\approx -2\sigma(T)/L$,
which justifies scaling and the linear behaviour of the solvation force slightly
on the liquid side of the pseudo-coexistence line.
Contribution to the solvation force arising
from the $L$-dependent regular  terms in the free energy  manifests itself strongly for $p=2$ 
and on the gas side of the pseudo-coexistence line
for all values of the decay exponent $p$.
There, except in the vicinity of the pseudo-capillary critical point for $p\ne 2$,
scaling does not hold and the solvation force is weak.

Close to the bulk critical temperature  the solvation force changes its behaviour significantly, i.e., 
the jump of $f_{solv}$ transforms into a minimum,  but the  effects of (pseudo) capillary condensations 
remain strong. This manifests itself at small values of  $|H|^{8/15}L$ where 
we observe data collapse for all values of $p$ and where the scaling function varies as $\approx -(|H|^{8/15}L)^{2}$, 
which implies $f_{solv} \sim H$ for small $H$ and fixed $L$
corresponding to the residual metastable bulk phase. 

For $p=2$ and 3, along the pseudo-coexistence line on its  gas side   and  along the isotherms   for large values of 
$|H|^{8/15}L$,  the regular contribution  (\ref{eq:1}) 
dominates the behaviour of  the solvation force. Because  $m^*(T)h_1<0$, at these thermodynamic points $f_{solv}$
is attractive.
On the liquid side of the pseudo-coexistence line and far away from (pseudo) capillary condensation and bulk
criticality, $f_{solv}$ becomes repulsive in agreement with (\ref{eq:1}) (see also Ref.~\cite{mdb}).
Near the bulk critical point and close to the capillary condensation (pseudo) transition the solvation force is 
attractive. In the case of short-range-like  boundary potential $p=50$, $f_{solv}$ is always attractive.

\acknowledgments 
We dedicate this paper to Bob Evans on the occasion of his 65 birthday for being always a source
of inspiration for us. 

We thank S. Dietrich for discussions. Numerical calculations were performed in
WCSS Wroc{\l}aw (Poland, grant 82).

\end{document}